\documentclass[a4paper,11pt]{article}
\usepackage{aaskaiid}
\usepackage{orcidlink}

\usepackage[normalem]{ulem}

\title{A Census of Variable and Transient Radio Sources Within High-Energy Neutrino Fields}
\ShortTitle{A Census of Radio Sources Within Neutrino Fields}

\author[1]{F. R\"osch \orcidlink{0009-0000-4620-2458}}
\ShortName{R\"osch et al.} 
\author[1]{M. Kadler}
\author[2, 3]{R. P. Deane}
\author[4]{P. G. Edwards \orcidlink{0000-0002-8186-4753}}
\author[1]{K. Mannheim}
\author[2, 5]{J. F. Radcliffe}

\affiliation[1]{Julius-Maximilians-Universit{\"a}t W{\"u}rzburg, Fakult{\"a}t f{\"u}r Physik und Astronomie, Institut f{\"u}r Theoretische Physik und Astrophysik, Lehrstuhl f{\"u}r Astronomie, Emil-Fischer-Str. 31, D-97074 W{\"u}rzburg, Germany}
\emailAdd{florian.roesch@uni-wuerzburg.de}
\affiliation[2]{Department of Physics, University of Pretoria, Lynnwood Rd, Hatfield, Pretoria, 0002, South Africa}
\affiliation[3]{University of the Witwatersrand, Enoch Sontonga Ave, Johannesburg, South Africa}
\affiliation[4]{CSIRO Space and Astronomy, PO Box 76, Epping, NSW 1710, Australia}
\affiliation[5]{Jodrell Bank Centre for Astrophysics, University of Manchester, Oxford Road, Manchester M13 9PL, UK}

\abstract{The origin of high-energy cosmic neutrinos detected by the IceCube observatory is a hotly debated topic in astroparticle physics. Neutrinos can be produced via interactions of high-energy protons with photons. There are multiple candidate source classes which can accelerate cosmic particles to the energies required to emit high-energy cosmic neutrinos and which have in common that they lead to variable/transient radio emissions. However, so far only a few active galactic nuclei (AGNs) could be associated with high confidence with IceCube neutrinos. The bulk of the diffuse 
neutrino flux might be emitted from a rather faint and numerous source population. The sub-mJy low-frequency radio sky may harbor these neutrino emitters that have gone unnoticed in previous searches. SKA-Mid continuum observations of high-energy neutrino fields will yield the most complete census of coincident transient and variable radio sources that might be associated with the neutrino emission. In previous MeerKAT observations of the IceCube gold alert IC240929A at 815 MHz, we detect about 550 faint radio sources inside the 90\% uncertainty region of this neutrino field with flux densities between $169\,\mathrm{\mu Jy}$ and $140\,\mathrm{mJy}$. 
In its AA4 configuration, SKA-Mid will achieve about four times the sensitivity of MeerKAT, allowing for the detection of even fainter radio sources within such neutrino fields. By adding wide-field VLBI analysis of the fields under consideration, all neutrino-candidate radio sources can be tested for high brightness-temperature compact emission (indicative of an AGN classification) and milliarcsecond-scale resolved structures.}

\begin{document}
\maketitle

\section{Introduction}
\label{sec:intro}
The origin of high-energy astrophysical neutrinos detected by the IceCube and, as recently reported \citep{KM3Net2025}, by the KM3NeT neutrino observatories is a hotly debated topic in astroparticle physics. They are believed to be produced via interactions of high-energy protons with other particles or with surrounding photon fields. Specifically, in proton-photon interactions pions are produced which subsequently decay into high-energy photons and neutrinos \citep[e.g.,][]{Mannheim1992,Abbasi2023a}. There are several source classes that can accelerate cosmic particles to the very high energies required to produce high-energy neutrinos and which have in common that they lead to variable and transient radio emissions. Active galactic nuclei (AGNs) have been particularly widely discussed and especially their subclass of blazars, radio-loud AGNs hosting relativistic jets pointed close to our line of sight, has been associated with some neutrinos in various studies \citep[e.g.,][]{IceCube2022,Buson2022,Plavin2023,Plavin2021,Plavin2020,Hovatta2021}. However, it has long been demonstrated that known $\gamma$-ray blazars can contribute only a small fraction to the observed diffuse neutrino flux \citep[e.g.,][]{Aartsen2017}. Furthermore, recent sample studies \citep[e.g.,][]{Abbasi2026,Abbasi2023b,Kouch2024} show no significant correlation between blazars and neutrino emission. Therefore, the bulk of the diffuse IceCube neutrino flux might be emitted from a rather faint and numerous source population, that may be harbored in the sub-mJy low-frequency radio sky and have gone unnoticed in previous searches. Other possible classes of high-energy neutrino emitting sources, besides blazars, include gamma ray bursts (GRBs), supernovae (SNe) and tidal disruption events (TDEs), which can all be associated with detectable radio transients.

Future SKA-Mid observations of high-energy neutrino fields will be able to detect such faint and variable radio sources. These SKA-Mid continuum observations will yield the most complete census of transient and variable radio sources inside the 90\% localization region of high-energy neutrino events.

\section{SKA-Mid Observations of High-Energy Neutrino Fields}
\label{sec:ska-mid}
IceCube sends out realtime automatic GCN Notice AMON alerts\footnote{\url{https://gcn.gsfc.nasa.gov/amon_icecube_gold_bronze_events.html}} to enable multiwavelengths follow-up observations of high-energy neutrino detections. These alerts report the directions of the neutrinos and their angular uncertainties and are classified as gold and bronze events with a signalness of $\geq 50\,\%$ and $\geq 30\,\%$, respectively, which is a measure for the probability of a neutrino event being of astrophysical origin. A few hours later a second, updated GCN Circular alert\footnote{\url{https://gcn.nasa.gov/circulars}} using a more sophisticated reconstruction algorithm is sent out \citep[][]{Abbasi2023b}. In the future, the KM3NeT observatory, which is mostly sensitive to neutrino emission from the Southern sky, will send out similar alerts \citep{Cecchini2025X3}. 

Triggering SKA-Mid observations on high-energy neutrino alerts with signalnesses of $\geq 50\,\%$ and 90\% localization regions that can be fully covered by the SKA-Mid field of view (FoV), will lead to the most complete census of transient and variable radio sources inside these neutrino regions. For SKA-Mid observations at band 1 or band 2, the FoV will be large enough to cover the entire 90\% localization region of the triggered neutrino fields. However, since some possible classes of high-energy neutrino emitting sources can be detected more easily and at earlier times at higher frequencies (see Sect.~\ref{sec:grbs} and Sect.~\ref{sec:sne} for more details), mosaicing the neutrino localization regions at higher frequencies would be an alternative option. 

Several consecutive observations of each triggered neutrino field will allow us to study the variability of the sources inside the 90\% uncertainty region. The first observation should be performed as close as possible to the neutrino event. After this, at least two additional epochs should be observed about two weeks and two months after the neutrino event, respectively, to study the variability on timescales of weeks to months. With such observations, the following source classes can be detected:

\subsection{Active Galactic Nuclei (AGNs) and Blazars}
\label{sec:blazars}
Although recent sample studies \citep[e.g.,][]{Abbasi2026,Abbasi2023b,Kouch2024} find no significant correlation between blazars and neutrino emitters, there are still a few individual AGNs that are associated with high-energy neutrino emission. In 2022, the \citet{IceCube2022} reported a correlation of neutrinos with  known $\gamma$-ray emitters at a significance of $3.3\,\sigma$ with the largest contributions from the nearby active galaxy NGC\,1068 and the three blazars TXS\,0506+056, PKS\,1424+240 and GB6\,J1542+6129. Another example for such a blazar association is PKS\,1424-418 that was associated at $2\sigma$ significance with the so-called BigBird neutrino event by \citet{Kadler2016}. Previous flux density monitoring studies \citep[e.g.,][]{Hovatta2021} found an association between radio flares of neutrino-candidate AGN and IceCube neutrino events. Such correlations are also observed by the TELAMON \citep{Eppel2024} and TANAMI/ATCA \citep{Stevens2012} monitoring programs using the Effelsberg 100-m telescope and the Australia Compact Telescope Array (ATCA) to study the radio spectra and flaring activity of the observed sources between 5 and 40\,GHz. The TELAMON and TANAMI/ATCA light curves of neutrino-candidate blazars show significant flaring activity on time scales of weeks to months \citep[e.g.,][]{Kadler2022, Eppel2023}. Furthermore, these sources show mostly flat and sometimes inverted spectra with spectral indices\footnote{$S(\nu)\propto\nu^\alpha$, where $S$ is the flux density, $\nu$ is the frequency, and $\alpha$ is the spectral index.} between $-0.5$ and $0.5$ \citep[e.g.,][]{Kadler2022} and high brightness temperatures \citep[e.g.,][]{Roesch2024}. Therefore, bright persistent but variable sources with flux densities $> 10\,\mathrm{mJy}$ and flat or inverted spectra are likely blazars, which can clearly be tested by very long baseline interferometry (VLBI) observations including SKA-Mid as phased array \citep[also see Sect.~\ref{sec:vlbi} and][]{Kadler2025}.

\subsection{Gamma Ray Bursts (GRBs)}
\label{sec:grbs}
Long duration GRBs are believed to originate from relativistic jets launched by black holes or magnetars. Within these jets, cosmic particles can be accelerated to sufficiently-high energies that neutrino production can result via proton-photon interactions \citep{Kurahashi2022}. On the other hand, short GRBs are thought to be caused by binary neutron star mergers. Until 2015, $58\,\%$ of these short GRBs were observed in the radio, but only $7\,\%$ of these sources were detected \citep{Fong2015}. Similar detection rates were also found by \citet{Chandra2012} by studying radio observations of a sample of 304 GRBs. They report a detection rate of $31\%$, including 2 out of 33 short GRBs. However, recent detections of a series of GRBs with the VLA at 6, 10 and 15 GHz \citep{Giarratana2023, Giarratana2024g, Giarratana2024i, Giarratana2024a, Giarratana2024b, Giarratana2024c, Giarratana2024d, Giarratana2024f, Giarratana2025c, Giarratana2025d, Giarratana2025a, Giarratana2025b, Giarratana2026b} indicate a consistently larger detection rate, showing that increasing sensitivity and tailored observation strategies can greatly improve the access to GRBs.

Although long duration GRBs generally last only a few seconds, they also show radio afterglows that are observable on timescales of minutes to weeks after the first detection. \citet{Weiler2002} studied the radio light curves of six GRB afterglows showing relatively faint flux densities at a few $\mu$Jy and peak values at a few mJy. More recent VLA observations at 6, 10 and 15 GHz of GRB afterglows taken about one week up to one month after the first detection measured flux densities of several tens to several hundreds of $\mathrm{\mu Jy}$ \citep{Giarratana2024h, Giarratana2024e, Giarratana2026a}. With its superior sensitivity at low frequencies, the SKA-Mid has the unique capability to detect these faint radio sources. However, the radio light curves of many GRB afterglows peak at a later time for lower frequencies than for higher ones \citep{Zhang2022}. With SKA-Mid observations at lower frequencies in band 1 or band 2, which cover the complete 90\% localization region of the observed neutrino events, we expect a higher flux density in later epochs for GRB-candidate variable sources. On the other hand, using a mosaicing approach with SKA-Mid observations at higher frequencies would allow potential GRBs within the neutrino field to be detected more easily in earlier epochs. Furthermore, GRB afterglows show a self-absorbed spectrum, with a spectral index of 2.5, below the peak frequency \citep{Laskar2023}. Therefore, radio transients that show increasing flux densities and a strongly inverted self-absorbed spectrum are likely GRB afterglows.

\subsection{Supernovae (SNe)}
\label{sec:sne}
SNe are often preceded by massive eruptions of circumstellar material or stellar envelope inflation. Such a system is considered a promising source of high-energy neutrinos \citep{Kurahashi2022}. \citet{Weiler2002} also studied the radio light curves of six SNe at frequencies from 22.5 GHz to 1.5 GHz. Similar to GRBs, the SN light curves also show relatively faint flux densities at a few $\mu$Jy with peaks at a few 100 mJy shifted to later times from the explosion for lower frequencies. Furthermore, typical SN light curves show a rapid increase in the radio flux density, that occurs later at lower frequencies, followed by a slower decline. For the six SNe studied by \citet{Weiler2002}, the rapid increase of the flux density at 1.5 GHz occurs at around 10 to 500 days after the explosion starting at flux densities of a few $\mu$Jy. Therefore, the SKA-Mid with its superior sensitivity, will be the perfect instrument to detect SNe at low frequencies. However, we expect to detect possible SNe inside the observed neutrino field only in the later epochs if the field is observed at band 1 or band 2. In contrast, if the neutrino field is observed at higher frequencies using a mosaicing approach, possible SNe within the neutrino field could likely be detected more easily already in the earlier epochs. Many SNe can not be detected by optical surveys, either because they are optically dim or obscured by large gas or dust clouds in their host galaxies \citep{Horiuchi2011}. Moreover, SNe which occur in the vicinity of the sun could be missed in optical surveys if they can only be observed after they have already become too faint. However, such SNe can be detected in the radio regime. SN2008iz, for example, was discovered in radio observations while it was not visible in the optical \citep{Horiuchi2011}.

\subsection{Tidal Disruption Events (TDEs)}
\label{sec:tdes}
TDEs occur when a star is disrupted by a supermassive black hole. According to \citet{Alexander2020}, most of those TDEs are radio quiet, with radio emission orders of magnitude fainter than that of radio-loud ones and a radio luminosity of $\nu L_\mathrm{\nu} < 10^{40}\,\mathrm{erg\,s}^{-1}$, while a small fraction produces radio-luminous, mildly relativistic jets. Within these jets, cosmic particles are accelerated to high energies which then results in high-energy neutrino production by proton-photon interactions, similar as in the jets of GRBs and radio-loud AGNs. Therefore, such jetted TDEs are thought to be high-energy neutrino emitters \citep{Kurahashi2022}. The radio detected TDE AT2019dsg has been suggested as a likely association to the IceCube neutrino event IC191001A \citep{Stein2021}. In this case, the neutrino event was observed 150 days after the peak of the optical luminosity \citep{Kurahashi2022}. This TDE was also observed at 1.3 GHz with MeerKAT showing an increase in the flux density with time at a flux density level of $\sim 100\,\mathrm{\mu Jy}$ \citep{Stein2021}. With its superior sensitivity, the SKA-Mid will be the perfect instrument to detect such faint radio sources. Observing at band 1 or band 2, or using the mosaicing approach at higher frequencies, we expect to detect an increase in the flux density at a level of $\sim 100\,\mathrm{\mu Jy}$ or below for possible TDEs in the observed neutrino fields.

\section{Pilot-Study Observations of the IceCube Neutrino Event IC240929A with MeerKAT}
\label{sec:meerkat}
On 29 September 2024, IceCube reported on the track-like neutrino event IC240929A, located at the J2000 coordinates R.A. = 180.66 deg (with 90\% containment uncertainties of $+0.57$ deg and $-0.71$ deg) and Dec. = 18.92 deg (with 90\% containment uncertainties of $+0.55$ deg and $-0.54$ deg) (GCN Circular \#37625\footnote{\url{https://gcn.nasa.gov/circulars/events/icecube-240929a}}). This event was initially reported via an automatic GCN Notice AMON GOLD alert\footnote{\url{https://gcn.gsfc.nasa.gov/notices_amon_g_b/139912_46959751.amon}} with a signalness of $71.17\%$.
As a pilot-study for future SKA-Mid observations of such neutrino fields, we observed this IceCube neutrino event with MeerKAT at $815\,\mathrm{MHz}$ twice, to perform a census of transient and variable radio sources inside its 90\% localization region. 
\begin{figure}[t]
    \centering
	\includegraphics[width=\columnwidth]{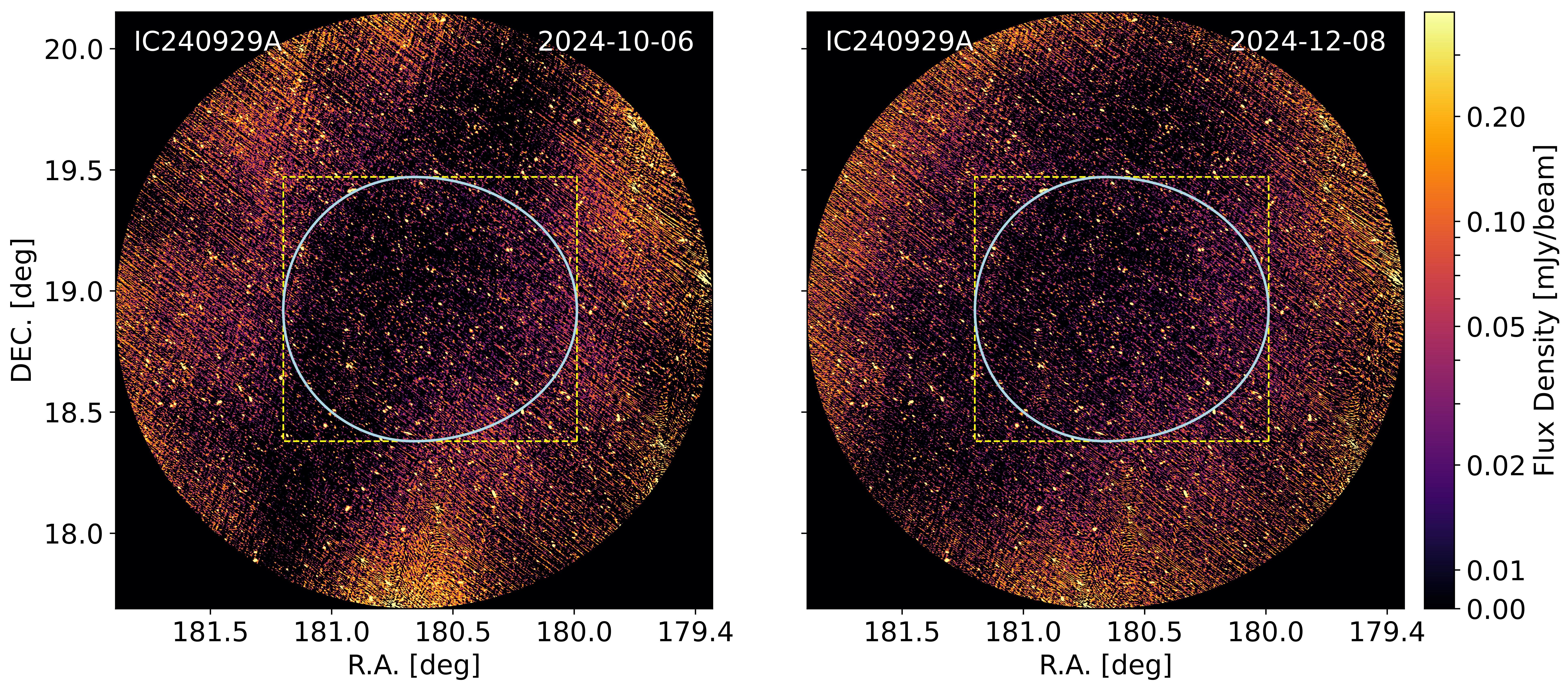}
    \caption{SDP pipeline images of MeerKAT observations at $815\,\mathrm{MHz}$ of the IceCube neutrino event IC240929A reported on 29 September 2024 in GCN Circular \#37625. The yellow dashed lines illustrates the 90\% containment box given by the 90\% containment uncertainties in R.A. and Dec., while the blue line shows the elliptically shaped 90\% localization region of the neutrino field.}
    \label{fig:images}
\end{figure}
To study the flux densities of the sources inside the neutrino field as close as possible to the neutrino event, the first observation took place seven days after the neutrino event on 6 October 2024. The second observation was performed about two months later on 8 December 2024, to study the variability of the detected radio sources. Both observations were performed with a duration of 1 hour on the neutrino field, leading to median rms noise values of $\sigma_\mathrm{rms, 1} = 51\,\mathrm{\mu Jy / beam}$ and $\sigma_\mathrm{rms, 2} = 54\,\mathrm{\mu Jy / beam}$ for the first and second observation. The automatically generated Science Data Processor (SDP) pipeline images provided by SARAO are plotted in Fig.~\ref{fig:images}. We used these two SDP pipeline images to search for variable radio sources inside the neutrino field.

\begin{figure}[t]
    \centering
	\includegraphics[width=\columnwidth]{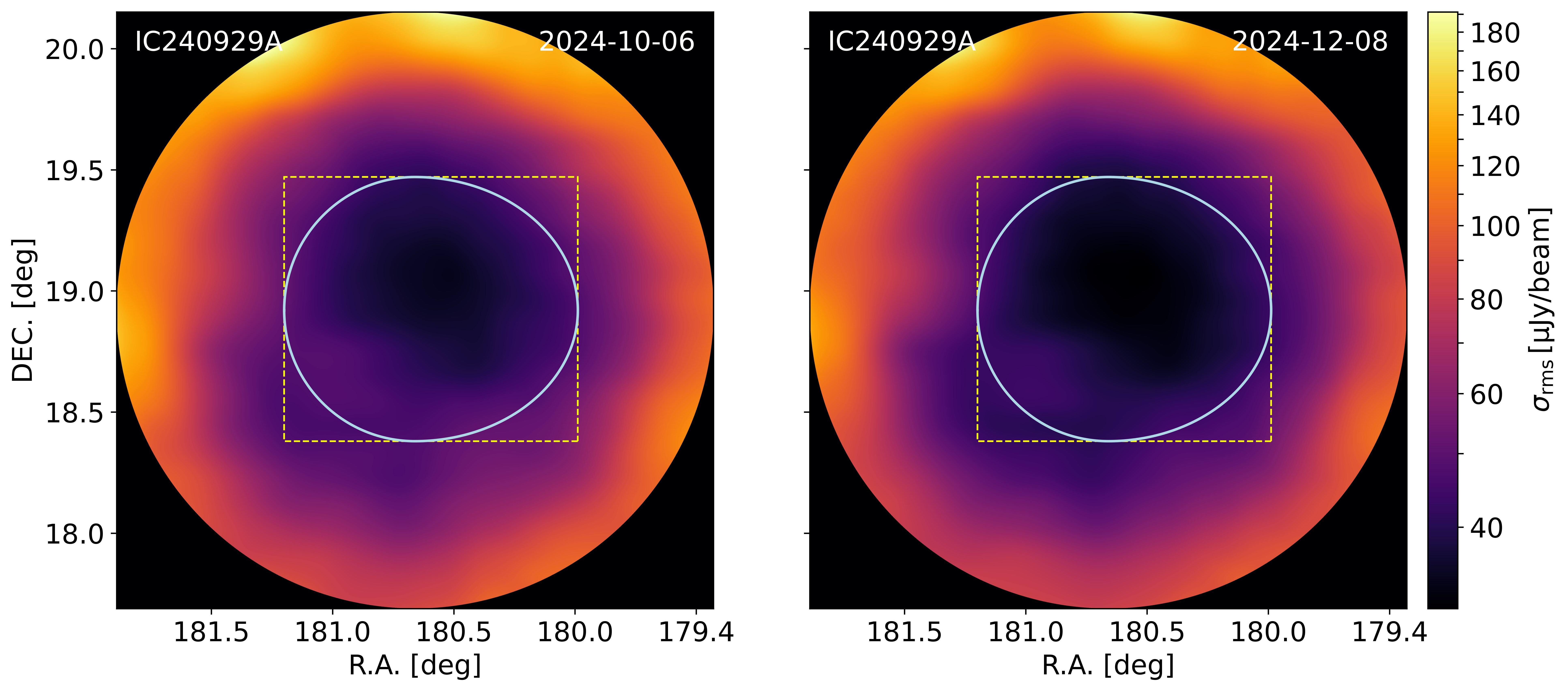}
    \caption{Estimated rms noise of MeerKAT observations at $815\,\mathrm{MHz}$ of the IceCube neutrino event IC240929A reported on 29 September 2024 in GCN Circular \#37625. The yellow dashed lines illustrates the 90\% containment box given by the 90\% containment uncertainties in R.A. and Dec., while the blue line shows the elliptically shaped 90\% localization region of the neutrino field. The best sensitivity with rms values of $\sim 31\,\mathrm{\mu Jy / beam}$ is achieved at the center of the neutrino field.}
    \label{fig:rms}
\end{figure}

Since the two SDP pipeline images have slightly different pixel sizes, we first re-gridded them on similar grids using the Python-package reproject \citep{reproject}. Furthermore, to ensure comparability we restored both epochs with a common beam, calculated as the smallest possible beam that includes both beams of both epochs, using the Python-package radio\_beam \citep{radio_beam}. To identify the radio sources in the re-gridded and restored images, we used the program PyBDSF \citep{pybdsf}. First, rms and mean images were produced for both observations by calculating the $3\sigma$-clipped rms and mean inside sliding boxes of size $623\times 623\,\mathrm{pixels}$, which were moved across the images in steps of $208\,\mathrm{pixels}$. These box and step sizes were calculated internally by PyBDSF. The resulting rms maps for both observations are shown in Fig.~\ref{fig:rms}. 
It can be seen that the sensitivities of the two SDP pipeline images achieve rms values of $\sigma_\mathrm{rms}\sim 31\,\mathrm{\mu Jy/beam}$ at the center of the observed neutrino field and increase to $\sigma_\mathrm{rms}\sim 190\,\mathrm{\mu Jy/beam}$ towards the edges of the FoV. These rms and mean images were then used to identify islands whose pixels have flux densities higher than $3\sigma_\mathrm{rms}$ above the mean. After this, Gaussian components were fitted to the islands so that all components have peak flux densities higher than $5\sigma_\mathrm{rms}$ above the mean. Finally, Gaussian components within an island were grouped into sources. More detailed information on the source fitting process can be found in the PyBDSF Documentation\footnote{\url{https://pybdsf.readthedocs.io/en/latest/index.html}}.

With this method, we find 1664 sources with flux densities between $196\,\mathrm{\mu Jy}$ and $912\,\mathrm{mJy}$ in the first epoch. While 684 of those sources with flux densities between $196\,\mathrm{\mu Jy}$ and $140\,\mathrm{mJy}$ are located in the 90\% containment box given by the 90\% containment uncertainties in R.A. and Dec. (yellow dashed lines in Fig.~\ref{fig:images}), 562 of these sources (with flux densities in the same range) lie inside the elliptically shaped 90\% localization region (blue line in Fig.~\ref{fig:images}) of the neutrino field. In the second epoch, we detect 1604 sources with flux densities between $81\,\mathrm{\mu Jy}$ and $426\,\mathrm{mJy}$. 662 of those are located in the 90\% containment box showing flux densities between $236\,\mathrm{\mu Jy}$ and $144\,\mathrm{mJy}$, while 552 sources lie in the ellipse-like region. In its AA4 configuration, the SKA-Mid will achieve a sensitivity about four times better than that of MeerKAT, allowing the detection of even fainter radio sources within similar neutrino fields.

\begin{figure}[h]
    \centering
	\includegraphics[width=0.9\columnwidth]{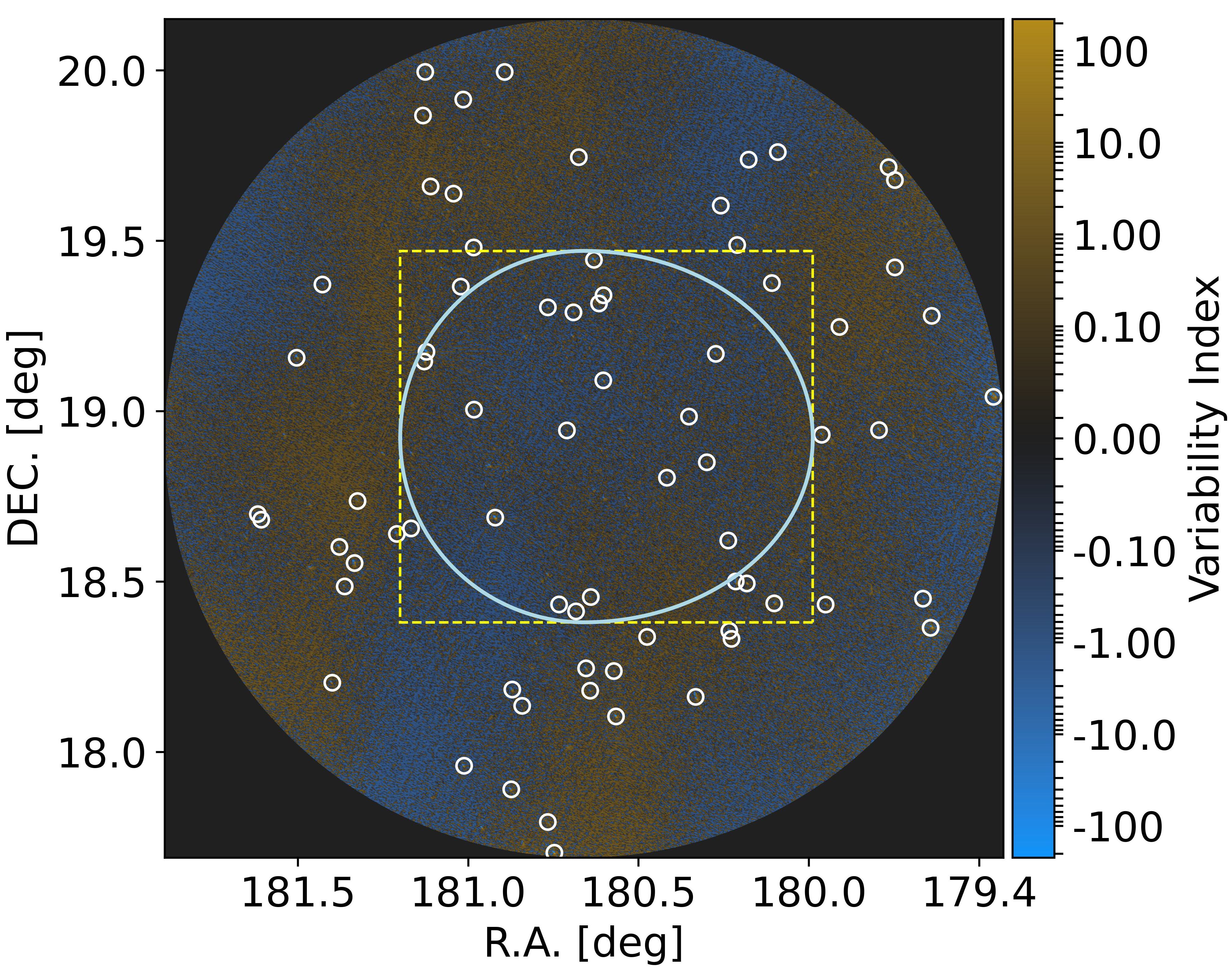}
    \caption{Variability index map as computed by equation~(\ref{eq:variability}). Negative variability indices denote an increase in flux density in the second epoch, while positive variability indices denote decreasing flux densities in the second epoch. Regions with significant variability indices that lie beyond the $3\sigma$ confidence interval of $-5.5 \leq V \leq 6.1$ are indicated by white circles. The yellow dashed lines illustrates the 90\% containment box given by the 90\% containment uncertainties in R.A. and Dec., while the blue line shows the elliptically shaped 90\% localization region of the neutrino field.}
    \label{fig:variability}
\end{figure}

To identify variable sources and transients inside the 90\% uncertainty region of the observed neutrino field, assuming that the radio interferometric noise is Gaussian distributed \citep[e.g.,][]{Condon1998}, we calculated the pixel-wise variability index $V$ using the following formula \citep[see also][]{Mooley2016, Radcliffe2019}
\begin{equation}
    V = \frac{\Delta S}{\sigma} = \frac{S_1 - S_2}{\sqrt{\sigma_\mathrm{rms, 1}^2 + \sigma_\mathrm{rms, 2}^2}},
	\label{eq:variability}
\end{equation}
where $S_\mathrm{i}$ is the flux density measured in the i-th epoch and $\sigma_\mathrm{rms, i}$ is its rms uncertainty taken from the rms map computed with PyBDSF. The derived variability index map is shown in  
Fig.~\ref{fig:variability}.

\begin{figure}[h]
    \centering
	\includegraphics[width=0.48\columnwidth]{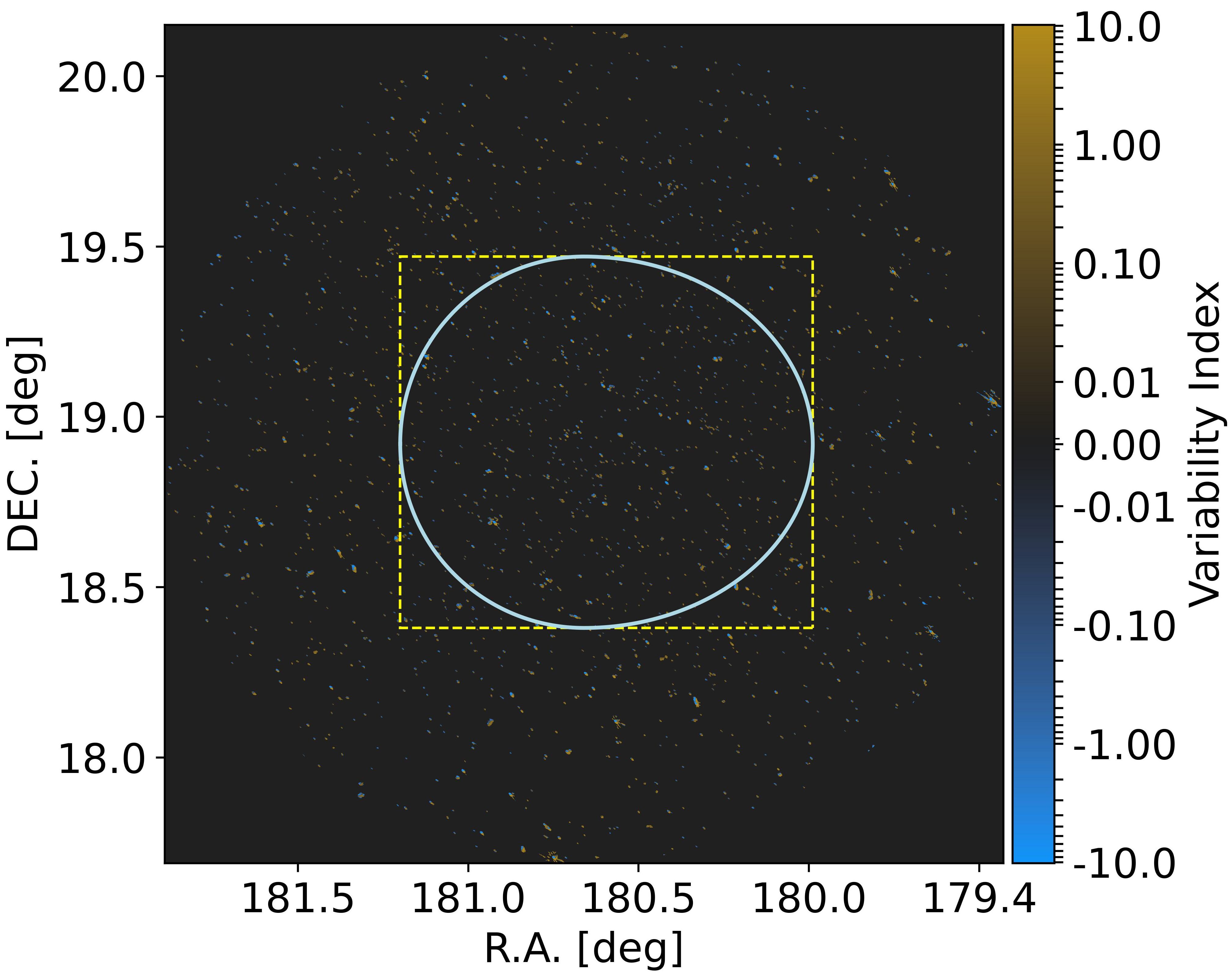}
    \includegraphics[width=0.49\columnwidth]{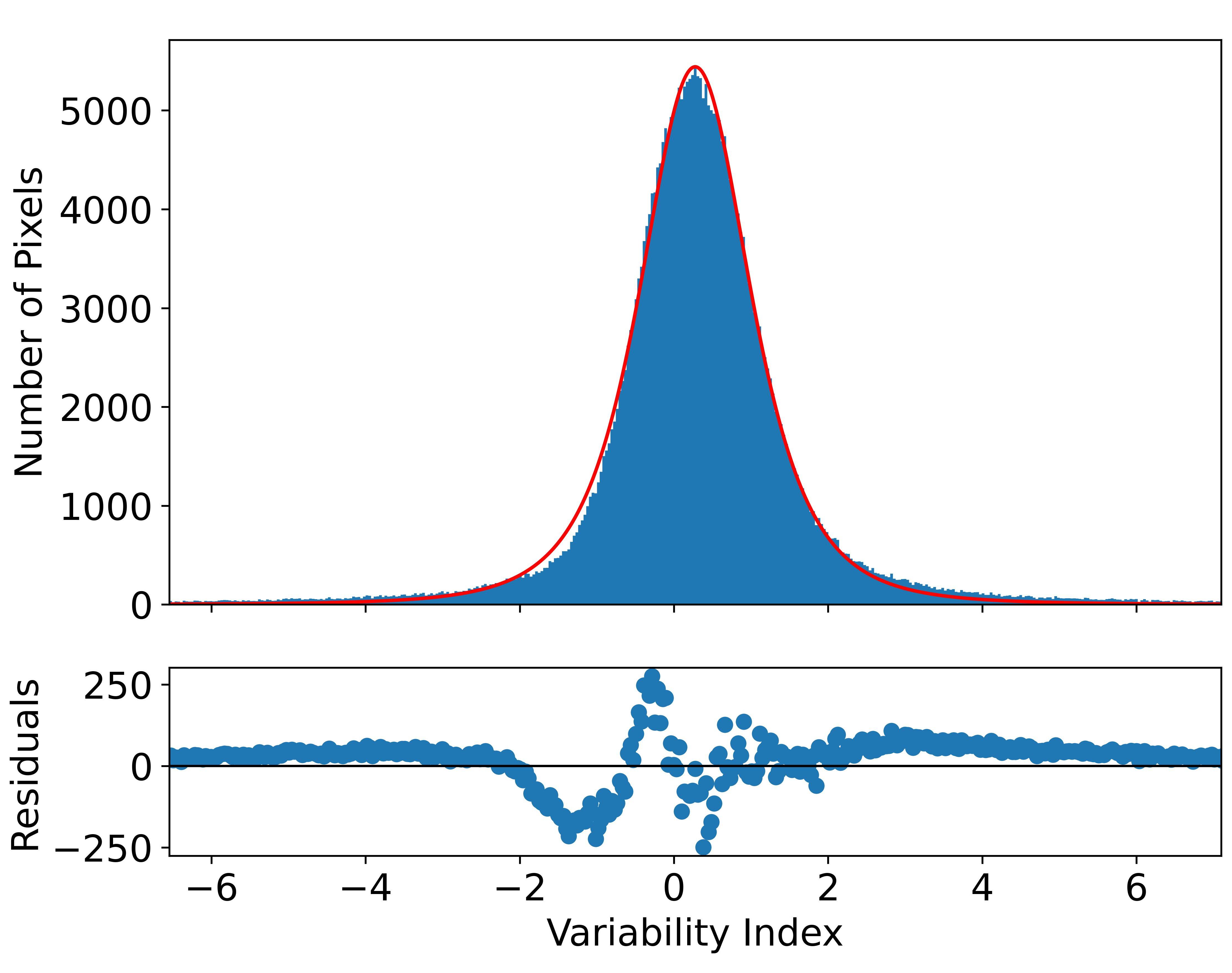}
    \caption{Left panel: Variability index map as computed by equation~(\ref{eq:variability}) showing only variability indices of island pixels that correspond to detected sources. Pixels that do not correspond to detected sources are set to zero. Negative variability indices denote an increase in flux density in the second epoch, while positive variability indices denote decreasing flux densities in the second epoch. The yellow dashed lines illustrates the 90\% containment box given by the 90\% containment uncertainties in R.A. and Dec., while the blue line shows the elliptically shaped 90\% localization region of the neutrino field. Right panel: Variability index distribution of island pixels that belong to detected radio sources fitted by a Student's t-distribution (red line). Pixels with variability indices outside of the two-sided Gaussian-equivalent $3\sigma$ confidence interval given by $-5.5 \leq V \leq 6.1$ are considered to be significant variable. These pixels are indicated by white circles in Fig.~\ref{fig:variability}. Note that the fact that the $3\sigma$ confidence interval is not symmetric around zero can probably be explained by systematics in the calibration.}
    \label{fig:sig_variability}
\end{figure}

The variability index $V$ is expected to be distributed according to the two-sided Student's t-distribution \citep[e.g.,][]{Bevington2003}. Therefore, we fitted the distribution of variability indices of island pixels that belong to detected sources with a Student's t-distribution to calculate its two-sided Gaussian-equivalent $3\sigma$ confidence interval (see Fig.~\ref{fig:sig_variability}). We derive a $3\sigma$ confidence interval of $-5.5 \leq V \leq 6.1$, symmetric around the mean value of $\bar V = 0.27584\pm 0.00046$.  
The fact that we find a significantly positive-offset $\bar V$ indicates that either the flux densities in the first epoch are too high, those in the second epoch are too low, or a combination of both. This is also supported by the residuals shown in the right panel of Fig.~\ref{fig:sig_variability}. These reveal a  lack of negative variability indices in the range $-2 \lesssim V \lesssim -0.5$ as well as a slight excess of positive variability indices in the range $2\lesssim V \lesssim 6$. Since we are using the SDP pipeline images, this can probably be explained by systematics in the calibration. Future improved imaging and self-calibration of the MeerKAT data may reduce the deviations from the expected distribution. 

We define a pixel as being significant variable if its variability index lies beyond the derived $3\sigma$ confidence interval of $-5.5 \leq V \leq 6.1$. Whenever a pixel that corresponds to the coordinates of a detected source was found to be significant variable, the corresponding source was defined as a variable or transient radio source. With this method, we find 72 variable radio sources within the MeerKAT FoV (indicated by white circles in Fig.~\ref{fig:variability}), 25 and 19 of which are located inside the 90\% uncertainty box and ellipse of the neutrino field, respectively. At the moment, it is difficult to clearly identify an outstanding source from the 19 variable sources as the potential counterpart of the neutrino event. Therefore, we will further investigate these sources also including information on their spectral indices.

\section{Wide-Field SKA-Mid VLBI Observations}
\label{sec:vlbi}
By adding wide-field SKA-Mid VLBI analysis to the above described observations, the detected neutrino-candidate radio sources can be tested for high brightness-temperatures and milliarcsecond-scale resolved structures. This is particularly helpful in the case of an AGN classification. AGNs associated with high-energy TeV and neutrino emission, observed with the Southern-hemisphere Long Baseline Array (LBA) within the TANAMI program \citep{Ojha2010} show high brightness-temperatures between $10^7\,\mathrm{K}$ and $10^{12}\,\mathrm{K}$ \citep{Benke2024,Roesch2024}. Furthermore, given the high sensitivity of the SKA-Mid, VLBI observations, including phased SKA-Mid into the array, will help to study the transverse structure of blazar jets and to test spine-sheath neutrino-production models. \citet{Tavecchioo2014} presented a model in which the high-energy protons of the fast inner spine of the blazar jet interact with soft target photons of the sheath, a slower jet-layer surrounding the spine. This leads to the production of pions which decay, producing both high-energy photons and neutrinos. This neutrino-production model leads to a limb-brightened jet structure which can be detected in VLBI images. The inclusion of phased SKA-Mid into existing VLBI arrays, like the LBA or the European VLBI Network (EVN), will improve the image sensitivity by a factor of $\gtrsim 4$. Therefore, by using wide-field SKA-Mid VLBI methods, we will be able to study possible limb-brightened structures of the blazar jets inside the observed neutrino fields \citep[also see][]{Kadler2025}.
Furthermore, such VLBI observations can also be used to study any other peculiar structures of the observed neutrino-candidate sources in general. Such structures could hint towards a possible neutrino association if they differ significantly from the structures of the other sources detected inside the neutrino field. Using several consecutive wide-field SKA-Mid VLBI observations, it would also be possible to investigate the structural evolution of the observed sources, such as the ejection of new components in blazar jets, to constrain their Lorentz factors and viewing angles \citep[e.g.,][]{Sumida2022, Eppel2026}, which can then be used to model the spectral energy distributions of the target sources \citep[e.g.,][]{Migliori2014}. More detailed information on SKA-Mid VLBI observations of blazar jets, GRBs, and TDEs can be found in \citet{Kadler2025}, \citet{Giarratana2025}, and \citet{Shu2025}, respectively.

\section*{Acknowledgements}
We thank the anonymous referee for their helpful comments and suggestions which improved the manuscript.
We acknowledge support from the Deutsche Forschungsgemeinschaft (DFG, grants 434448349 and 443220636 [FOR5195: Relativistic Jets in Active Galaxies]). 
The MeerKAT telescope is operated by the South African Radio Astronomy Observatory, which is a facility of the National Research Foundation, an agency of the Department of Science, Technology and Innovation.

\bibliographystyle{abbrvnat-maxbibnames4}
\bibliography{chapter}

\end{document}